\documentclass[%
 reprint,
 superscriptaddress,
showpacs,
preprintnumbers,
 amsmath,amssymb,
 aps,prl,
floatfix,
]{revtex4-1}

\usepackage{graphicx}
\usepackage{dcolumn}
\usepackage{bm}
\usepackage{todonotes}
\usepackage[mathlines]{lineno}
\usepackage{wrapfig}
\usepackage{rotating}
\usepackage{epstopdf}
\usepackage{multirow}

\usepackage{silence}
\begin{document}

\title{Spectroscopy of $^{50}$Sc and {\it ab initio} calculations of {\it B(M3)} strengths}

\author{A.B.~Garnsworthy}
 \email{garns@triumf.ca}
 \affiliation{TRIUMF, 4004 Wesbrook Mall, Vancouver, BC, V6T 2A3, Canada}
 
\author{M.~Bowry} 
\affiliation{TRIUMF, 4004 Wesbrook Mall, Vancouver, BC, V6T 2A3, Canada}

\author{B.~Olaizola} 
\affiliation{TRIUMF, 4004 Wesbrook Mall, Vancouver, BC, V6T 2A3, Canada}

\author{J.D.~Holt} 
\affiliation{TRIUMF, 4004 Wesbrook Mall, Vancouver, BC, V6T 2A3, Canada}

\author{S.R.~Stroberg} 
\affiliation{TRIUMF, 4004 Wesbrook Mall, Vancouver, BC, V6T 2A3, Canada}

\author{S.~Cruz} 
\affiliation{Department of Physics and Astronomy, University of British Columbia, Vancouver, BC V6T 1Z4, Canada}
\affiliation{TRIUMF, 4004 Wesbrook Mall, Vancouver, BC, V6T 2A3, Canada}

\author{S.~Georges} 
\affiliation{TRIUMF, 4004 Wesbrook Mall, Vancouver, BC, V6T 2A3, Canada}

\author{G.~Hackman}
\affiliation{TRIUMF, 4004 Wesbrook Mall, Vancouver, BC, V6T 2A3, Canada}

\author{A.D.~MacLean}
\affiliation{Department of Physics, University of Guelph, Guelph, ON, N1G 2W1, Canada}

\author{J.~Measures} 
\affiliation{TRIUMF, 4004 Wesbrook Mall, Vancouver, BC, V6T 2A3, Canada}

\author{H.P.~Patel} 
\affiliation{TRIUMF, 4004 Wesbrook Mall, Vancouver, BC, V6T 2A3, Canada}

\author{C.J.~Pearson} 
\affiliation{TRIUMF, 4004 Wesbrook Mall, Vancouver, BC, V6T 2A3, Canada}

\author{C.E.~Svensson}
\affiliation{Department of Physics, University of Guelph, Guelph, ON, N1G 2W1, Canada}

\date{\today}

\begin{abstract}
The GRIFFIN spectrometer at TRIUMF-ISAC has been used to study excited states and transitions in $^{50}$Sc following the $\beta$-decay of $^{50}$Ca. Branching ratios were determined from the measured $\gamma$-ray intensities, and angular correlations of $\gamma$ rays have been used to firmly assign the spins of excited states. The presence of an isomeric state that decays by an $M3$ transition with a $B(M3)$ strength of 13.6(7)\,W.u. has been confirmed. We compare with the first {\it ab initio} calculations of $B(M3$) strengths in light and medium-mass nuclei from the valence-space in-medium similarity renormalization group approach, using consistently derived effective Hamiltonians and $M3$ operator. The experimental data are well reproduced for isoscalar $M3$ transitions when using bare $g$-factors, but the strength of isovector $M3$ transitions are found to be underestimated by an order of magnitude.
\end{abstract}

\maketitle

{\it Introduction -}
Electromagnetic transitions between nuclear states carry away energy and angular momentum from the nucleus to obtain a more stable arrangement of the constituent nucleons.
Magnetic octupole ($M3$) transitions represent a change of 3$\hbar$ of angular momentum with no change in parity between the initial and final nuclear state.
$M3$ decay transitions are rarely observed in nuclei, as the deexcitation is usually dominated by lower-order electromagnetic decays, specifically magnetic dipole ($M1$) and electric quadrupole ($E2$).
Indeed, in the few situations where the lowest-order transition allowed by angular momentum conservation is $M3$, the nuclear state usually has a half life of milliseconds to hours, therefore surviving long enough that there is competition from $\beta$ decay. 
This situation makes them excellent examples of spin-trap isomeric nuclear states~\cite{Walker1999}. 

The calculation of transition strengths is a particularly sensitive test of theory, as it relies on a good reproduction of both initial- and final-state wavefunctions, as well as a realistic description of the transition operator. 
Electroweak transitions probe additional physics that is not sampled in the usual calculations of ground-state and excitation energies because the various operator structures will be sensitive to different components of the wavefunction.
The systematics of $M3$ transition strengths, which are expected to be dominated by a change between maximum and minimum orbital angular momentum couplings, have the potential to provide additional insights over the more common $M1$ and $E2$. While the rather exotic physics of $M3$ transitions was explored thoroughly within a phenomenological context by Brown {\it et al.}~\cite{Brown1980}, there has been little discussion in terms of more microscopic studies since.

The development of a first-principles, or {\it ab initio}, description of atomic nuclei is a central challenge in nuclear theory. 
The task is complicated because the exact form of nuclear interactions is not known, and there is great complexity in solving the nuclear many-body problem.
Progress on the former has been made via chiral effective field theory (EFT) \cite{Epel09RMP,Mach11PR} and the similarity renormalization group (SRG) \cite{Bogn07SRG,Bogn10PPNP}, which allow for a systematic and consistent expansion of nuclear forces, where three-nucleon (3N) interactions have emerged as an essential component \cite{Hebe15ARNPS}. While promising, there is currently no established procedure for constraining or optimizing the free parameters of chiral EFT. Thus a number of different interactions, relying on different strategies have been produced recently, ranging from those incorporating data from medium-mass systems \cite{Ekst15sat}, to local interactions appropriate for quantum monte carlo calculations \cite{Geze14local}, to including explicit delta degrees of freedom \cite{Piar16del,Ekst17del}.
On the many-body side, developments of {\it ab initio} techniques continue a rapid push from light- to medium-mass systems \cite{Carl15RMP,Barr13PPNP,Hage14RPP,Herg16PR,Dickhoff2004,Roth09ImTr}. 
In particular, the valence-space formulation of the in-medium similarity renormalization group (IMSRG)  \cite{Tsuk12SM,Bogn14SM,Stro17ENO,Stro16TNO} has been established as a powerful approach that extends the reach of {\it ab initio} many-body theory to essentially all open-shell nuclei at least to the tin region.

Until very recently, {\it ab initio} calculations of electroweak transitions were not possible beyond the lightest nuclei but are now accessible with both coupled-cluster theory \cite{Ekst14GT2bc} and the IMSRG \cite{Parz17Trans}.
In this article we present new experimental data that confirms the existence of a $M3$ transition in $^{50}$Sc, bringing the number of these transitions identified to six in nuclei up to $A=$50. 
Motivated by this new result, we have performed the first calculations of $B(M3$) strengths using the {\it ab initio} valence-space \mbox{(VS-)IMSRG}, with consistent effective valence-space Hamiltonians and $M3$ operators, and present them here.

{\it Experimental Details -}
The isotope $^{50}$Ca (T$_{1/2}$ = 13.9(6)\,s \cite{Warburton1970}) was produced from reactions induced in a 22.49\,g/cm$^2$ Ta target by a 500\,MeV proton beam delivered by the TRIUMF Cyclotron \cite{Bylinskii2013}. The position of the 60\,$\mu$A proton beam on the ISOL target was continuously rastered. This was the first time beam rastering was employed for delivering radioactive beam to an experimental station at ISAC and allowed for a tighter proton beam spot resulting in a higher localized power density in the Ta target. The calcium atoms created in the target that diffused out of the material were ionized using resonant-laser ionization and accelerated to 20\,keV, mass separated and delivered to the experimental station. The typical beam intensity of $^{50}$Ca was $\sim 10^6$\,ions/s. A small amount of surface-ionized $^{50}$K ($T_{1/2}$=472(4)~ms \cite{Langevin1983}) was also present in the beam.

The ions were stopped in a mylar tape at the central focus of the Gamma-Ray Infrastructure For Fundamental Investigations of Nuclei (GRIFFIN) spectrometer \cite{Svensson2013,Rizwan2016,Garnsworthy2017}. GRIFFIN consists of an array of 16 high-purity Germanium (HPGe) clover detectors coupled to a series of ancillary detectors. Fifteen HPGe clovers were used in the present work. An array of plastic scintillator paddles (SCEPTAR) was used for the detection of $\beta$ particles. Four 5.1\,cm diameter and 5.1\,cm deep cylindrical lanthanum bromide (LaBr$_3$(Ce)) scintillators with a 5\% cerium doping were used for $\gamma$-ray fast timing. The GRIFFIN clovers were positioned at a source-to-detector distance of 11~cm from the implantation point whereas the LaBr$_3$(Ce) detectors were at 12.5\,cm. A 20\,mm thick delrin plastic absorber shell was placed around the vacuum chamber to prevent $\beta$ particles from reaching the HPGe detectors while minimizing the flux of Bremsstrahlung photons created as the $\beta$ particles were brought to rest.

The experiment ran as a series of cycles with two time structures employed. The two sets of cycles included a period of background measurement (0.5/3.5\,s), source accumulation (3/5\,s), source decay (3/40\,s), and source removal (1.5/1.5\,s). This cycling allowed the periodic removal of the long-lived $^{50}$Sc daughter (T$_{1/2}$ = 102.5(5)\,s) activity from sight of the detectors. Data was collected in the shorter cycle mode for 56~mins, and the longer cycle mode for 51\,mins.

Energy and timing signals were collected from each detector using the GRIFFIN digital data acquisition system \cite{Garnsworthy2017}, operated in a triggerless mode. In addition the signals from the LaBr$_3$(Ce) detectors were used as input to a set of NIM analogue electronics for fast coincident timing. An Ortec 935 constant-fraction discriminator for each detector fed a set of logic modules that ultimately present the start and stop signals to a set of Ortec 566 time-to-amplitude converter NIM modules for which the output is digitized in a GRIF-16 digitizer. HPGe energy and efficiency were calibrated using standard radioactive sources of $^{133}$Ba, $^{152}$Eu, $^{60}$Co and $^{56}$Co with the necessary corrections for coincidence summing applied.

\begin{figure}[ht]
	\centering
    \includegraphics[width=\linewidth]{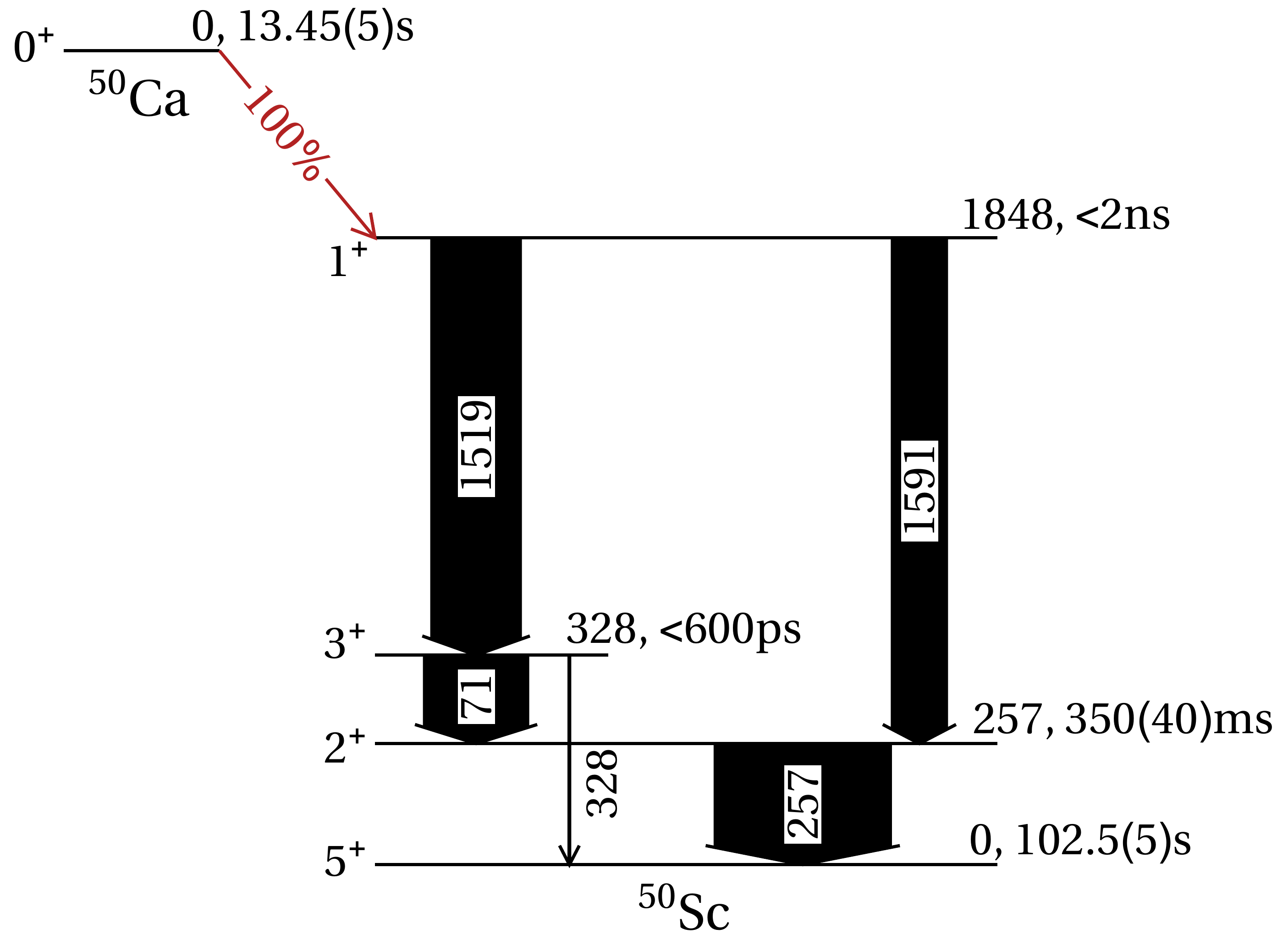}
    \caption{\label{fig:scheme}Level scheme of the levels observed in $^{50}$Sc following the $\beta$ decay of $^{50}$Ca. The width of the arrows represent the relative total intensity of the transition measured in this work. The positioning of the levels have been modified for easier visualization.}
\end{figure}

{\it Experimental Results -}
The $^{50}$Sc level scheme of states and transitions observed in the decay of $^{50}$Ca are shown in Figure \ref{fig:scheme}. A more precise measurement of the $^{50}$Ca ground state of 13.45(5)\,s has been made by fitting the time distribution of the 1519 and 1591\,keV $\gamma$ rays and applying the same analysis methods described in Ref. \cite{Dunlop2016}.
The efficiency-corrected relative intensities of the $\gamma$ rays emitted from $^{50}$Sc were determined from the $\gamma$-ray singles spectrum and are presented in Table \ref{Tab:TranStrengths}. The total internal conversion coefficients are calculated using BrIcc \cite{Kibedi2008}.
The spin and parity of the state at 1848\,keV has been previously assigned as $1^+$ from a measured $L=0$ transfer in a ($\alpha$,d) reaction \cite{Fister1994} and the log $ft$=4.1(2) value from the 0$^+$ ground state of the $^{50}$Ca $\beta$ decay parent \cite{Alburger1984}.

\begin{figure}[ht]
	\centering
    \includegraphics[width=0.9\linewidth]{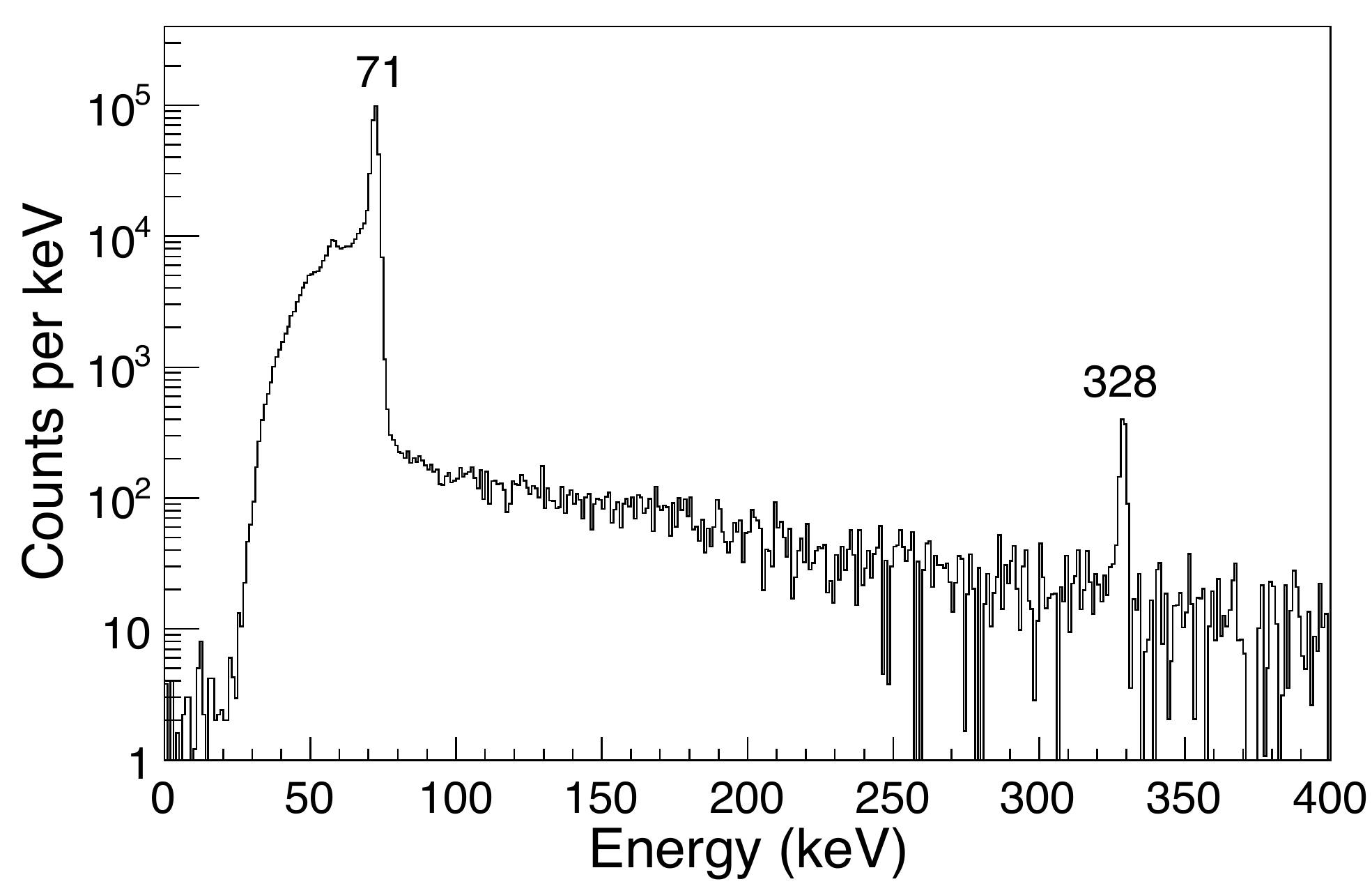}
    \caption{\label{fig:spectrum}Gamma-ray energy spectrum in coincidence with the 1519\,keV transition. The 328\,keV branch is observed for the first time with a $\gamma$-ray branching ratio of 0.78(8)\%.}
\end{figure}

The results of our work do not reproduce the discrepancy in the intensity balance around the 328\,keV state that was reported by Alburger {\it et al.} \cite{Alburger1984}.
The transition from the 328\,keV state to the ground state has been observed in this work for the first time as can be seen in Figure \ref{fig:spectrum}. The total branching ratio of 0.75(8)\% is consistent with the previously suggested upper limit of $<$0.68\% but is now a measurement.
We reduce the upper limit on the $\beta$ decay branching ratio of the 257\,keV state from $<$2.5\% to $<$1\% from an examination of the observed gamma-ray intensities in $^{50}$Sc and $^{50}$Ti.

\begin{table*}[ht]
\begin{centering}
\caption{\label{Tab:TranStrengths}Spectoscopic information for $^{50}$Sc. Experimental and theoretical transition strengths are shown in Weisskopf units. Theoretical transition strengths are calculated from standard operators using effective charges ($e_\pi=1.5$, $e_\nu=0.5$) for electric transitions and bare $g$-factors for the magnetic transitions from wavefunctions produced with the KB3G $pf$ shell interaction and the VS-IMSRG. The half life value of the 257\,keV state is taken from Ref. \cite{Alburger1984} whereas the others are from the current work. Internal conversion coefficients, $\alpha_{Tot}$, are from Ref. \cite{Kibedi2008}.}
\begin{tabular}{ccc|cccc|cccc}
\hline
\multirow{2}{*}{Trans.} & $E_{Exp}$ & Mult. & $T_{1/2}$ & $I_\gamma$ & $\alpha_{Tot}$ & $I_{Tot}$ & Exp. & Exp. & $pf$-KB3G & VS-IMSRG \\
 & (keV) & & & (This work) &  & & (Lit.) & (This work) & Wavefunctions & Wavefunctions \\
\hline
$2^+ \rightarrow 5^+$ & 257 & $M3$ & 350(40)~ms & 100(2) & 0.022 & 102(2) & 13.3(16) & 13.6(7) & 13.9 & 11.1\\
$3^+ \rightarrow 2^+$ & 71 & $M1$ & $<$600~ps & 58.4(60) & 0.039 & 60.7(62) &  $>$0.01 & $>$0.2 & 2.8 & 3.5\\
                      & 71 & $E2$ &  &  &  &  & $>$390 & $>$11 & 2.8 & 4.2\\
$3^+ \rightarrow 5^+$ & 328 & $E2$ & $<$600~ps & 0.46(5) & 0.003 & 0.46(5) & $>$0.002 & $>$0.2 & 2.0 & 2.4\\
$1^+ \rightarrow 3^+$ & 1519 & $E2$ & $<$2~ns & 59.6(17) & 1.4x10$^{-4}$ & 59.6(17) & $>$0.0004 & $>$0.002& 3.3 & 3.9\\
$1^+ \rightarrow 2^+$ & 1591 & $M1$ & $<$2~ns & 36.3(10) & 1.3x10$^{-4}$ & 36.3(10) & $>$2x10$^{-7}$ & $>$1x10$^{-6}$ & 0.5 & 0.1\\
\hline
\end{tabular}
\end{centering}
\end{table*}%

In order to explain the intensity imbalance reported previously, an ($E2/M1$) mixing ratio of $\delta$=0.40(15) was adopted by the evaluator \cite{NDS_A50}. The consequence of this mixing ratio corrects the intensity imbalance through the larger internal conversion coefficient of the $E2$ multipole in comparison to $M1$. 
However, this significant contribution of $E2$ implies an unreasonably large $B(E2)$ value (as was noted by the authors of \cite{Alburger1984}). 

\begin{figure}[ht]
	\centering
    \includegraphics[width=0.95\linewidth]{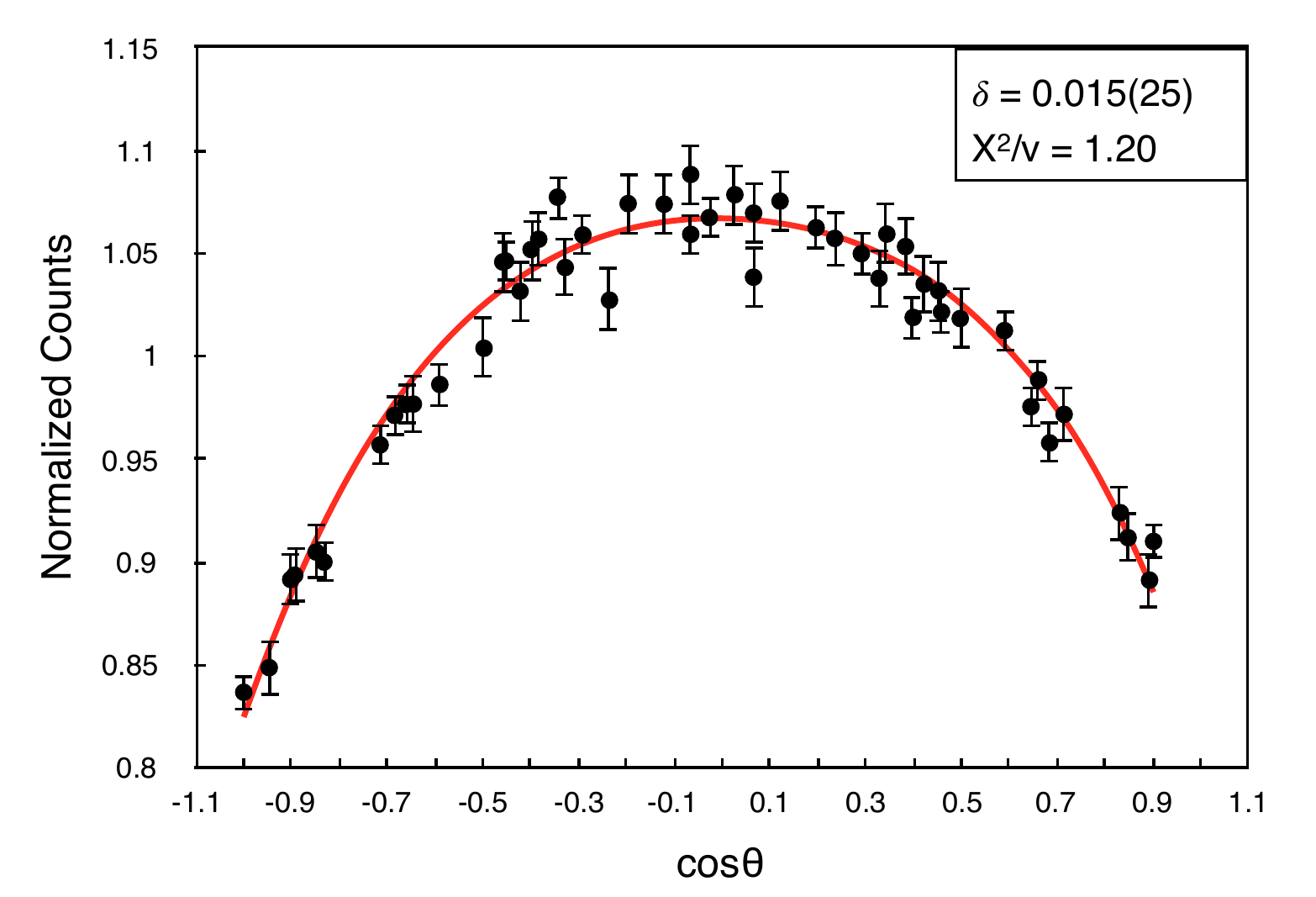}
    \caption{\label{fig:Ang_corr}The $\gamma-\gamma$ angular correlation of the 1519-71\,keV cascade indicating an (E2/M1) mixing ratio of +0.015(25) for the 71\,keV transition, and firmly assigning the spins of the 257 and 328\,keV states as $2^+$ and $3^+$ respectively.}
\end{figure}

In the present work we have directly measured the ($E2/M1$) mixing ratio of the 71~keV transition using $\gamma-\gamma$ angular correlations. 
The GRIFFIN spectrometer offers 51 unique angles for this analysis (the zero degree opening angle case is omitted) and the angular correlation for the 1519-71\,keV $\gamma-\gamma$ cascade is shown in Figure \ref{fig:Ang_corr}. The data point for relative detector angles of 18.8$^{\circ}$, which corresponds to neighboring crystals in the the same clover, was excluded from this analysis due to overlap with Compton-scattered events of the 1591\,keV $\gamma$ ray. The data were compared to a series of GEANT4 simulations that investigated different spin combinations and values of the mixing ratios for the transitions. Each GEANT4 simulation, containing $10^9$ events,  accounts for the geometric and finite solid-angle effects of the detectors.
The excellent statistics obtained in the measurement for this cascade ($>$400,000 coincidences) allowed for a precise value of $\delta$=+0.015(25) to be determined.
The GEANT4 simulated $\gamma-\gamma$ angular correlation with this mixing ratio value fitted the data with a reduced chi-squared of 1.20. The spins of the 328 and 257\,keV states are now firmly assigned as $J=3$ and 2 respectively. This confirms the transition between the 257\,keV 2$^+$ state and 5$^+$ ground state to be of $M3$ multipolarity. The possibility that this is a hindered $E2$ transition is excluded. It was not possible to make a new measurement of the half life of this isomer in this work so the value of 350(40)\,ms is used from Ref. \cite{Alburger1984}.

Upper limits of $<$10\,ns had been assigned to the half lives of the 328 and 1848\,keV states from $\beta-\gamma$ coincidence timing \cite{Alburger1984}. 
The half life of the 1848\,keV state was investigated using the generalized centroid difference method \cite{Centroid}. 
This method is sensitive to energy-dependent time-walk effects and usually measurements are made relative to known lifetimes with gamma-ray energies that cover the energy range of interest. In this case independent measurements were made using $\beta-\gamma$ coincidences between a SCEPTAR paddle and either the 1519 or 1591\,keV $\gamma$ rays de-exciting the 1848\,keV state detected in a HPGe detector.
The 1553\,keV transition from the $2_1^+$ state in the daughter nucleus $^{50}$Ti~\cite{NDS_A50} was then used to extract the lifetime. The $\beta$ decay of $^{50}$Sc populates the $6_1^+$ state ($T_{1/2}$=418(17)\,ps) with I$_\beta=88.4(15) \, \%$ and the $4_1^+$ state ($T_{1/2}$=5.3(11)\,ps) with I$_\beta=8.4(18)\, \%$, so these two half-lives are carried into the $2_1^+ \rightarrow 0^+_1$ transition. This effect was taken into account during the analysis.
A conservative upper limit of $<2$\,ns is assigned to the 1848\,keV state based on this analysis.

The time difference between the 71 and 1519\,keV gamma rays detected in the LaBr$_3$(Ce) detectors in the present study was examined. No lifetime component of the 328\,keV state was discernible from the prompt response. Following a thorough investigation of the systematic effects in this non-optimized experimental setup a conservative upper limit of $<600$\,ps is assigned to the 328\,keV state from this work.

The new data firmly establish the multipolarity of all observed transitions following the $\beta$ decay of $^{50}$Ca and allow more stringent experimental limits to be placed on the transition strengths.

\begin{table}[t]
\begin{centering}
\caption{\label{Tab:Occupations}Single-particle-orbital occupation numbers of the wavefunctions calculated with the $pf$-shell KB3G (lower value) and VS-IMSRG (upper value) interactions.}
\begin{tabular}{c|cccc|cccc}
\hline
State & \multicolumn{4}{c}{Proton} & \multicolumn{4}{c}{Neutron}\\
  & $0f_{7/2}$ & $1p_{3/2}$ & $0f_{5/2}$ & $1p_{1/2}$ & $0f_{7/2}$ & $1p_{3/2}$ & $0f_{5/2}$ & $1p_{1/2}$\\
\hline
\multirow{2}{*}{5$^+$} & 0.97 & 0.01 & 0.01 & 0.00 & 7.62 & 1.12 & 0.20 & 0.06\\ 
  & 0.98 & 0.01 & 0.01 & 0.00 & 7.70 & 1.08 & 0.17 & 0.04\\ 
  &  &  &  &  &  &  &  & \\
\multirow{2}{*}{2$^+$} & 0.95 & 0.03 & 0.02 & 0.01 & 7.63 & 1.07 & 0.24 & 0.06\\ 
 & 0.96 & 0.02 & 0.01 & 0.01 & 7.71 & 1.03 & 0.21 & 0.05\\ 
  &  &  &  &  &  &  &  & \\
\multirow{2}{*}{3$^+$} & 0.94 & 0.04 & 0.01 & 0.00 & 7.64 & 1.00 & 0.20 & 0.16\\ 
 & 0.97 & 0.02 & 0.01 & 0.00 & 7.71 & 0.95 & 0.18 & 0.16\\ 
  &  &  &  &  &  &  &  & \\
\multirow{2}{*}{1$^+$} & 0.65 & 0.19 & 0.09 & 0.06 & 7.72 & 0.38 & 0.78 & 0.12\\ 
 & 0.76 & 0.15 & 0.03 & 0.06 & 7.74 & 0.29 & 0.86 & 0.10\\  
\hline
\end{tabular}
\end{centering}
\end{table}%

{\it Calculations and Discussion -}
Shell model calculations were performed for $^{50}$Sc with the NuShellX@MSU shell-model code \cite{Brown2014} using the phenomenological KB3G interaction \cite{Poves2001} in the $pf$ valence space ($0f_{7/2}$, $1p_{3/2}$, $0f_{5/2}$, $1p_{1/2}$), known to well reproduce experimental data in this region.  In addition, we derive {\it ab initio} shell-model Hamiltonians within the VS-IMSRG framework \cite{Tsuk12SM,Bogn14SM,Stro17ENO,Stro16TNO}, based on two-nucleon (NN) and three-nucleon (3N) forces derived from chiral effective field theory \cite{Epel09RMP,Mach11PR}. The particular input NN+3N interaction, developed in Refs.~\cite{Hebe11fits,Simo16unc,Simo17SatFinNuc}, begins from a chiral NN interaction at next-to-next-to-next-to leading order (N$^3$LO) \cite{Ente03EMN3LO,Mach11PR} and is evolved with the free-space SRG \cite{Bogn07SRG} to a low-momentum scale $\lambda_{\mathrm{NN}}=1.8\,\mathrm{fm}^{-1}$.  Unconstrained couplings of the 3N force at order N$^2$LO are  fit to reproduce the triton binding energy and $\alpha$ particle charge radius at $\Lambda_{\mathrm{3N}}=2.0\,\mathrm{fm}^{-1}$.  This Hamiltonian, which is fit to only few body data, predicts realistic saturation properties of infinite symmetric nuclear matter \cite{Hebe11fits}, and also reproduces ground-state energies across the nuclear chart from the $p$ shell to the nickel region and beyond \cite{Hag16,Rui16,Simo17SatFinNuc,Lasc17Cd}.

Starting in a single-particle spherical harmonic oscillator (HO) basis with energy $\hbar\omega=16$\,MeV, we first transform the input Hamiltonian to the Hartree-Fock (HF) basis, then use the Magnus formulation of the VS-IMSRG \cite{Morr15Magnus,Herg16PR}, with the ensemble normal ordering procedure~\cite{Stro17ENO}, which captures the bulk effects of residual 3N forces among valence nucleons, to produce an approximate unitary transformation which decouples the $^{40}$Ca core.  A second transformation is performed to decouple a specific $pf$-shell valence-space Hamiltonian appropriate for $^{50}$Sc. These results are well converged within the basis size $e=2n+l \le e_{\mathrm{max}}=12$ and $e_1 + e_2 + e_3 \le E_{\mathrm{3max}} = 16 $.

We begin by comparing the KB3G and VS-IMSRG wavefunction composition for the lowest few states of $^{50}$Sc in the form of single-particle orbital occupations, shown in Table \ref{Tab:Occupations}. Here we see that the results of the two calculations are remarkably similar. The difference in occupation number is less than one tenth of a nucleon for all states and single-particle orbitals for both protons and neutrons.

Using the accepted set of effective charges ($e_\pi=1.5$, $e_\nu=0.5$) in this region \cite{Poves2001} and the bare spin, orbit and tensor $g$-factors ($g_{\pi s}$=5.586, $g_{\pi l}$=1.0, $g_{\pi p}$=0.0, $g_{\nu s}$=-3.826, $g_{\nu l}$=0.0, $g_{\nu p}$=0.0) for protons ($\pi$) and neutrons ($\nu$) for all multipolarities in both models, transition strengths between each state observed in $\beta$ decay are shown in comparison to the experimental results in Table \ref{Tab:TranStrengths}. This allows for a direct comparison between the wavefunctions calculated within the phenomenological and VS-IMSRG frameworks. A comparison with an effective $M3$ operator derived consistently within the VS-IMSRG framework is given later. Investigations of consistently-derived $M1$ and $E2$ operators can be found in Ref. \cite{Parz17Trans}.
Here we see that despite nearly identical occupation numbers, the $B(M1)$ and $B(E2)$ values between KB3G and VS-IMSRG can differ by up to 30\%, while the difference in $B(M3)$ is nearly 20\%, due to one-body transition density amplitudes resulting from the two valence-space Hamiltonians.  
Nonetheless, both calculations reproduce well the large newly measured $M3$ transition strength with the $g$-factors mentioned above.

\begin{table*}[ht]
\begin{centering}
\caption{\label{Tab:M3Exp}Experimentally known $M3$ transition strengths in nuclei up to $A$=50. Only transitions for which the lowest order allowed multipolarity is 3 are included. The experimental data for $\gamma$-ray energy ($E_\gamma$), level half life ($T_{1/2}$) and $\gamma$-ray intensity ($I_\gamma$) are taken from Refs. \cite{NSR1961Sc09,NSR1970Ch37,NSR1972Br53,NSR1980Jo11,NSR2011Ni18,NDS_A34,NSR1974Gr48,NSR1962KI09,NSR2008Le12} and the current work. The internal conversion coefficients ($\alpha_{Tot}$) are taken from Ref. \cite{Kibedi2008} in order to determine the total intensity ($I_{Tot}$) of each transition. The experimental and calculated $B(M3)$ values are expressed in Weisskopf units where 1\,W.u.=1.6501$A^{4/3}\mu_N^2fm^4$.}
\begin{tabular}{cccc|ccccc}
\hline
Isotope & $E_\gamma$ & $J_i^\pi\rightarrow J_f^\pi$ & $\Delta T$ & $T_{1/2}$ & $I_\gamma$ & $\alpha_{Tot}$ & $I_{Tot}$ & Exp. \\
& (keV) & & & & & & & $B(M3)$ \\
\hline
$^{24}$Na \cite{NSR1961Sc09,NSR1970Ch37,NSR1972Br53,NSR1980Jo11} & 472.2074(8) & 1$^+\rightarrow$4$^+$ & 0 & 20.18(10)ms & 0.9995(5) & 0.000469(7) & 0.9995(5) & 9.10(7) \\ 
$^{24}$Al \cite{NSR2011Ni18} & 425.8(1) & 4$^+\rightarrow$1$^+$& 0 & 131.3(25)ms & 0.83(3) & 0.001144(16) & 0.83(3) & 2.4(6) \\ 
$^{34}$Cl \cite{NDS_A34} & 146.36(3) & 3$^+\rightarrow$0$^+$& 1 & 31.99(3)min & 0.383(5) & 0.1656(24) &  0.446(6) &  0.10(1) \\
$^{38}$Cl \cite{NSR1972Br53,NSR1974Gr48,NSR1962KI09} & 671.365(8) & 5$^-\rightarrow$2$^-$& 0 & 715(3)ms & 0.3826(8) & 0.000599(9) & 1 &  0.0118(8) \\
$^{38}$K \cite{NSR2008Le12} & 130.1(2) & 0$^+\rightarrow$3$^+$& 1 & 924.33(27)ms & 8(1)$\times$10$^{-6}$ & 0.394(7) & 0.00033(4) &  0.29(10)\\ 
$^{50}$Sc \cite{Alburger1984} & 257.895(1) & 2$^+\rightarrow$5$^+$& 0 & 350(40)ms & 0.97(3) & 0.0350(5) & 0.99(1)  & 13.6(7) \\
\hline
\end{tabular}
\end{centering}
\end{table*}%

In addition to the properties of $^{50}$Sc, we have also examined the $B(M3)$ values for the other known cases of $M3$ transitions in $sd$- and $pf$-shell nuclei up to $A=50$ shown in Table \ref{Tab:M3Exp}. The experimental $B(M3)$ strengths for these five other cases have been calculated from the available literature data \cite{NSR1961Sc09,NSR1970Ch37,NSR1972Br53,NSR1980Jo11,NSR2011Ni18,NDS_A34,NSR1974Gr48,NSR1962KI09,NSR2008Le12}. It is important that the total branching ratio be used in the calculation of the $B(M3)$ strength as the internal conversion decay can be significant for high-multipolarity, low-energy transitions. We also note the sensitivity to the transition energy as the energy term for $M3$ is to the seventh power.

Table \ref{Tab:M3Theory} presents theoretical calculations of these $B(M3)$ transition strengths. For $^{24}$Na, $^{24}$Al, $^{34}$Cl and $^{38}$K we use the phenomenological USDB interaction \cite{Brow06USD} and a VS-IMSRG Hamiltonian derived in the standard $sd$ valence space specifically for each nucleus.
In the case of $^{38}$Cl we use the phenomenological SDPF-U interaction \cite{Nowa07SDPFU} and take a proton $sd$, neutron $pf$ space for the VS-IMSRG calculations. 
While bare $g$-factors are used in all cases, Brown {\it et al.} explored the ability of quenched spin $g$-factors to capture effects from core polarization as well as using a HF basis for $M3$ transitions \cite{Brown1980}.  While this has not been done here with the more modern USDB interaction, we can directly study the impact of both in the VS-IMSRG. Therefore strengths derived from operator matrix elements in both the HO and HF basis are shown in Table \ref{Tab:M3Theory}, the former being more comparable to the phenomenological shell model results and the latter being consistent with the VS-IMSRG wavefunctions. We note a systematic increase in the final transition rate when using the HF basis, except for in the case of $^{38}$Cl. In $^{24}$Na it is not clear what is driving the particularly large increase in the transition strength between a HO and HF basis.

\begin{table*}[ht]
\begin{centering}
\caption{\label{Tab:M3Theory}Comparison of calculations of $M3$ transition strengths in nuclei up to $A$=50. Details of the calculations are given in the text. The $B(M3)$ value is obtained by  $<A_{1b}+A_{2b}>^2 / (2J_i+1)$ in units of $\mu_N^2fm^4$, where $A_{1b}$ and $A_{2b}$ are the 1-body and 2-body amplitudes respectively. The experimental and calculated $B(M3)$ values are expressed here in Weisskopf units where 1\,W.u.=1.6501$A^{4/3}\mu_N^2fm^4$.}
\begin{tabular}{ccc|c|c|cc|ccc}
\hline
& & & & Phenomenological & \multicolumn{2}{c|}{VS-IMSRG Bare Op.} & \multicolumn{3}{c}{VS-IMSRG Effective Op.}\\
Isotope & $J_i^\pi\rightarrow J_f^\pi$ & $\Delta T$ & Exp. & shell model & HO & HF & & &  \\
& & & $B(M3)$ & $B(M3)$ & $B(M3)$ & $B(M3)$ & $A_{1b}$ & $A_{2b}$ & $B(M3)$ \\
\hline
$^{24}$Na & 1$^+\rightarrow$4$^+$ & 0 & 9.10(7) & 19.9 & 3.82 & 9.36 & 51.199 & -12.154 & 4.45 \\ 
$^{24}$Al & 4$^+\rightarrow$1$^+$& 0 & 2.4(6) & 2.72 & 1.99 & 2.86 & -50.545 & 8.026 & 1.76 \\ 
$^{34}$Cl & 3$^+\rightarrow$0$^+$& 1 &  0.10(1) & 0.157 & 0.017 & 0.019 & -3.791 & 5.072 & 0.0013\\
$^{38}$Cl & 5$^-\rightarrow$2$^-$& 0 &  0.0118(8) & 0.0003 & 0.010 & 0.0013 & 8.007 & -0.8648 & 0.022 \\
$^{38}$K & 0$^+\rightarrow$3$^+$& 1 &  0.29(10) & 0.324 & 0.011 & 0.021 & -1.962 & 3.752 & 0.015 \\ 
$^{50}$Sc & 2$^+\rightarrow$5$^+$& 0 & 13.6(7) & 13.9 & 11.14 & 15.03 & 12.008 & -0.824 & 9.62 \\
\hline
\end{tabular}
\end{centering}
\end{table*}%

Good agreement with experiment is found for the phenomenological approach except for $^{38}$Cl, which may be due to a lack of cross-shell neutron correlations allowed in the SDPF-U interaction. We also see that the VS-IMSRG in the HF basis reproduces quite well the data for isoscalar transitions when using the bare $M3$ transition operator, implying that the wavefunctions determined from this theory closely match those of the phenomenological approach. The isovector transitions in the odd-odd, $N=Z$ nuclei, $^{34}_{17}$Cl$_{17}$ and $^{38}_{19}$K$_{19}$, however, are underestimated by an order of magnitude. 

Finally we discuss the results using the VS-IMSRG effective $M3$ operator, calculated here for the first time.  
As discussed in \cite{Parz17Trans}, the same transformation that is used to acquire the valence-space Hamiltonian is applied to decouple an effective valence-space $M3$ operator, which includes two-body physics induced by the VS-IMSRG transformation.
For the nuclei in the $sd$ shell, we use an effective operator calculated with $e_{\mathrm{max}}=12$, while for the other cases we use $e_{\mathrm{max}}=10$, and always with $E_{\mathrm{3max}}=16$. Similar to $M1$ transitions \cite{Parz17Trans}, we would expect excitations mediated by the $M3$ operator vertex, including core polarization, to account partially for the missing physics captured in the phenomenological quenching factors \cite{Brown1980}.  The impact of using a valence-space effective operator consistent with the Hamiltonian is shown in the rightmost column of Table \ref{Tab:M3Theory}. We first note that there is no consistent reduction in the final $B(M3)$ value, since in $^{24}$Na the effective operator increases the result.  As well there is little consistency in the magnitude of the effect, which ranges from almost negligible in $^{38}$K to several orders of magnitude in $^{34}$Cl. However the calculations do reproduce the qualitative trend in $M3$ strengths below $A=50$, where in particular, we predict the new transition in $^{50}$Sc to be the largest among the known cases.

The underestimation of the isovector transitions is also seen in the calculation using the effective VS-IMSRG operator. This appears to originate in a suppression of the one-body transition amplitude from induced two-body components of the operator, as can be seen in the rightmost columns of Table \ref{Tab:M3Theory}. While there is always a cancellation between the one- and two-body parts, due to their opposite signs, for isovector transitions the magnitude of the two-body amplitudes is of the same size or larger than the one-body amplitudes. This provides a net suppression of around 60\% for isovector transitions, while for isoscaler transitions the two-body amplitude is never more than 20\%. This indicates a clear lack of many-body convergence for isovector transitions, the origin of which is unclear but will be studied further in the future.

Finally, it is also expected that neglected effects of meson-exchange currents should play a significant role in a proper description of electroweak currents in general. These can be derived consistently with the forces within chiral effective field theory. The effects of these currents have recently been shown to be appreciable for $M1$ transitions in light nuclei \cite{Past13momM12b} but have not yet been studied for heavier systems.  Furthermore, to our knowledge, there is no estimate of the importance of such physics in $M3$ transitions.  
The inclusion of such physics in the VS-IMSRG framework will be essential to draw a firm conclusion on the seemingly inconsistent effects of the valence space effective operator revealed by the current work.

{\it Conclusions:}
The GRIFFIN spectrometer at TRIUMF-ISAC has been used to study excited states and transitions in $^{50}$Sc following the $\beta$-decay of $^{50}$Ca. Branching ratios were determined from the measured $\gamma$-ray intensities. Angular correlations of $\gamma$ rays have been used to firmly assign the spins of excited states to confirm the existence of an isomeric state that decays by an $M3$ transition with a $B(M3)$ strength of 13.6(7)\,W.u, the strongest known $M3$ transition in the $A\leq 50$ mass region. 

We have performed calculations of $B(M3$) strengths in nuclei below $A=50$ using an {\it ab initio} approach with the VS-IMSRG. The experimental data are well reproduced for isoscalar $M3$ transitions when using bare $g$-factors, while the strength of isovector $M3$ transitions are found to be underestimated by an order of magnitude. 
We have calculated an effective valence-space $M3$ operator for the first time within the VS-IMSRG approach and find an inconsistent effect across the nuclei studied. Since the topic of effective operators is quite new in the {\it ab initio} community, there is not yet any intuition for the expected behavior of an effective $M3$ operator. Even $M1$ and $E2$ operators have been studied only very recently within the IMSRG \cite{Parz17Trans}, where $M1$ transition strengths were generally found to agree with experiment, up to expected effects from meson-exchange currents, while highly collective $E2$ transitions are significantly underpredicted with respect to experiment. Further exploration of neglected physics in $M3$ transitions is also needed, particularly the role of meson exchange currents, and the unclear many-body convergence for isovector transitions is needed to clarify the impact of effective $M3$ operators within {\it ab initio} methods.

{\it Acknowledgements:}
We would like to thank the operations and beam delivery staff at TRIUMF for providing the radioactive beam. We thank K. Hebeler, J. Simonis and A. Schwenk for providing the 3N matrix elements used in this work and for valuable discussions. This work was supported in part by the Natural Sciences and Engineering Research Council of Canada (NSERC). C.E.S. acknowledges support from the Canada Research Chairs program. The GRIFFIN spectrometer was jointly funded by the Canadian Foundation for Innovation (CFI), TRIUMF, and the University of Guelph.
TRIUMF receives federal funding via a contribution agreement through the National Research Council Canada (NRC).  Computations were performed with an allocation of computing resources at the J\"ulich Supercomputing Center (JURECA).

\bibliographystyle{apsrev}
\bibliography{50Sc}

\end{document}